\documentclass[11pt]{article}
\usepackage{graphics}
\usepackage{epsfig}

\def\beqra{\begin{eqnarray}} 
\def\eeqra{\end{eqnarray}}
\def\DN{\Delta N_\nu}
\def\be{\begin{equation}}  
\def\bea{\begin{eqnarray}}  
\def\ee{\end{equation}}     
\def\eea{\end{eqnarray}}     
\def\gsim{\mathrel{\raise.3ex\hbox{$>$\kern-.75em\lower1ex\hbox{$\sim$}}}}
\def\lsim{\mathrel{\raise.3ex\hbox{$<$\kern-.75em\lower1ex\hbox{$\sim$}}}}

\begin{document}

\vskip .7cm

\vspace{30cm}

\begin{center}
{\Large \bf  Structure Formation with Mirror Dark Matter: CMB and LSS}
\vskip .7cm
\begin{center} 
{\large Zurab Berezhiani $^{a,b,}$\footnote{E-mail:
     {\tt 
 berezhiani@aquila.infn.it}}}, ~
{\large Paolo Ciarcelluti $^{a,}$\footnote{E-mail:
     {\tt ciarcelluti@lngs.infn.it}}}, ~ 
{\large Denis Comelli $^{b,}$\footnote{E-mail:       {\tt  comelli@fe.infn.it}}}, \\ 
{\large Francesco L. Villante $^{b,c,}$\footnote{E-mail:
     {\tt villante@fe.infn.it}}} 
\end{center} 
{\it 
$^a$  Dipartimento di Fisica, Universit\`a di L'Aquila, 67010 Coppito  
AQ, and  \\  INFN, Laboratori Nazionali del Gran Sasso, 67010 Assergi AQ, 
Italy \\ 
$^b$  INFN, sezione di Ferrara, 44100 Ferrara, Italy\\
$^c$  Dipartimento di Fisica, Universit\`a di Ferrara, 41000 Ferrara, 
Italy \\
\vspace{8pt}
}
\end{center}
\vskip .3cm
\begin{abstract}
In the mirror world hypothesis the mirror baryonic component emerges
as a possible dark matter candidate. 
An immediate question arises: how the mirror
baryons behave and what are the differences from the more 
familiar dark matter candidates as e.g. cold dark matter?
In this paper we answer quantitatively to this question. 
First we discuss the dependence of the relevant scales for 
the structure formation (Jeans and Silk scales) on the two 
macroscopic parameters necessary to define the model: 
the temperature of the mirror plasma 
(limited by the Big Bang Nucleosynthesis)
and the amount of mirror baryonic matter. 
Then we perform a complete quantitative calculation 
of the implications of mirror dark matter on the cosmic microwave 
background and large scale structure power spectrum.
Finally, confronting with the present observational data, 
we obtain some bounds on the mirror parameter space.

\end{abstract}

\setcounter{footnote}{0}
\setcounter{page}{1}

\vspace{0.3cm}

%------------------------------------------------------------------------------------------------------------------------

\section{Introduction}

The idea that there may exist a hidden mirror sector of 
particles and interactions with exactly the same properties  
as that of our visible world was suggested 
long time ago \cite{defmirror}.  
The basic concept is to have a theory given by the 
product $G\times G'$ of two identical gauge factors with 
the identical particle contents, which could naturally emerge 
e.g. in the context of $E_8\times E'_8$ superstring.
(From now on the ``primed" $(')$ fields and quantities stand 
for the mirror (M) sector to distinguish from the ones belonging 
to the observable or ordinary (O) world.)
In particular, one can consider 
a minimal symmetry $G_{\rm SM}\times G'_{\rm SM}$  
where $G_{\rm SM}=SU(3)\times  SU(2)\times U(1)$ stands for the
standard model of observable particles: three families of 
quarks and leptons $q_i, ~u^c_i, ~d^c_i; ~l_i, ~e^c_i$
($i=1,2,3$) and the Higgs doublet $\phi$, 
while $G'_{\rm SM}=[SU(3)\times SU(2)\times U(1)]'$ is its mirror
gauge counterpart with analogous particle content:
fermions $q'_i, ~u'^c_i, ~d'^c_i; ~l'_i, ~e'^c_i$  and 
the Higgs $\phi'$. 
The M-particles are singlets of $G_{\rm SM}$ and vice versa,
the O-particles are singlets of $G'_{\rm SM}$.    
More generally, one can have in mind 
the grand unified extensions like $SU(5)\times SU(5)'$, 
$SO(10)\times SO(10)'$, etc. 
Besides the gravity, two sectors could communicate by other means  
\cite{IJMP}.  
In particular, ordinary and mirror neutral particles could mix:
e.g. photons \cite{Holdom} or neutrinos \cite{FV-BM}, 
two sectors could have a common gauge group of flavour \cite{PLB98}
or common Peccei-Quinn symmetry \cite{BGG}. 

A discrete symmetry $G\leftrightarrow G'$ interchanging
corresponding fields of $G$ and $G'$, so called mirror parity,  
guarantees that two particle sectors are described by 
identical Lagrangians, with all coupling constants 
(gauge, Yukawa, Higgs) having the same pattern. 
As a consequence the two sectors should have the same microphysics. 
In the particular case in which $G$ sector is left-handed 
and $G'$ sector is right-handed, this discrete symmetry 
can be interpreted as the true parity \cite{FV-parity}.

If the mirror sector exists, then the Universe
along with the ordinary photons, neutrinos, baryons, etc.
should contain their mirror partners. 
One would naively think that the O- and M-sectors, having 
the same microphysics, should have also the same 
cosmology.\footnote{ More generically, M-parity 
could be spontaneously broken and the weak scales
$\langle \phi \rangle$ and $\langle \phi' \rangle$
could be different, which would lead to somewhat different 
particle physics in the mirror sector \cite{BDG}. }  
Then one would expect that M-particles are present in the same
amount in the Universe as O-ones, which would be in immediate 
conflict with the Big Bang Nucleosynthesis (BBN) bounds 
on the effective number of extra light neutrinos $\DN$:   
the mirror photons, electrons and neutrinos would give 
a contribution equivalent to $\DN\simeq 6.14$.
However two sectors may have different initial conditions. 
In particular, after inflation, the two sectors could be   
reheated at different temperatures, $T'_R \neq T_R$, 
which can be achieved in certain inflationary 
models \cite{BDG}-\cite{z}. 
If the O- and M-particles interact weakly enough 
(which condition is automatically fulfilled  if the two worlds 
communicate only via the gravity), 
the two systems do not come into thermal equilibrium 
at later epoch and they evolve independently,
maintaining  approximately constant the ratio 
$x=T'/T$ among their temperatures. 
The BBN constraints are satisfied if $x$ is sufficiently low. 
Namely, the most conservative bound $\DN <1$ \cite{LSV}   
implies that  $x < 0.64$ \cite{z}.

The difference of the temperatures $T' < T$ breaks the symmetry 
between the cosmological properties of O- and M-sectors. 
Namely, the number density of M-photons is much smaller than 
that of O-ones, $n'_\gamma/n_\gamma = x^3 \ll 1$. 
It is important to understand, whether the mirror baryon
density $n'_b$ could be comparable or even larger than the ordinary 
one $n_b$, 
in which case they could constitute dark matter of the Universe, 
or at least its significant fraction. 

The cosmological evolution of the Mirror Universe 
was studied in ref. \cite{z} and all the key epochs  
(baryogenesis, nucleosynthesis, recombination etc.) were analyzed 
in details.  
It was shown, in particular, that in the context of the GUT or 
electroweak baryogenesis scenarios the condition 
$T'<T$ yields that $\eta'_b \geq \eta_b$, where 
$\eta_b = n_b/n_\gamma$ and $\eta'_b = n'_b/n'_\gamma$ 
respectively are the baryon asymmetries in O- and M-sectors \cite{z}. 
However, $\eta'_b/\eta_b\geq 1$ is not yet enough 
for having $\beta = n'_b/n_b \geq 1$. 
Since $\beta = x^3 (\eta'_b/\eta_b)$, the latter requires 
much stronger condition $\eta'_b/\eta_b \geq x^{-3}$.  
As it was demonstrated in ref. \cite{z}, this condition 
can be satisfied for a certain range of parameters, with   
the values of $x$ which are not too low, $x\geq 0.01$ or so. 
However, more appealing situation emerges in a leptogenesis 
scenario due to particle exchange between the 
ordinary and mirror sectors suggested in
ref. \cite{Bento}, which predicts $\beta \geq 1$ but also 
implies that $\beta < 10$ or so, 
and thus can explain naturally the near coincidence between 
the visible (O-baryon) matter density $\Omega_b$ and 
the dark (M-baryon) matter density $\Omega'_b$  
in a rather natural way \cite{IJMP}.   

If $\Omega'_b \geq \Omega_b$, 
mirror baryons emerge as a possible dark matter candidate;
they can contribute the dark matter of the Universe along with 
the cold dark matter or even constitute a dominant dark 
matter component. 
An immediate question arises: how the mirror
baryon dark matter (MBDM) behaves and what are the differences
from the more familiar dark matter candidate as the cold dark 
matter (CDM)? 

The peculiar properties of mirror dark matter were discussed
qualitatively in \cite{z}, and this analysis was confirmed and 
extended in refs. \cite{ignatiev,Paolo}. 
In this paper we complete this program
giving a complete quantitative presentation of the implications
of the MBDM for the cosmological large scale structure (LSS) 
and the cosmic microwave background (CMB) anisotropies. 

The plan of the paper is as follows. 
In the next section we analyze the relevant length scales for 
structure formation in the mirror photon-baryonic sector. 
In section 3 we describe the effect of the evolution of 
perturbations in linear regime and compute the power spectra for 
the LSS and CMB for various values of the two parameters $x$ and $\beta$.
The main differences with respect to a standard CDM scenario are 
discussed. Finally, our main conclusions are summarized in section 4.

\section{Relevant length scales}

Mirror matter may seem a tremendously complicated dark matter candidate.
However, from the point of view of structure formation, it can
be described relatively simply. The microphysics of the mirror sector
is in fact well defined, being identical to that of our sector. All the
differences with respect to the ordinary world can be described in terms
of two macroscopic parameters which are the only free parameters 
in the model:
\be \label{x-beta} 
x\equiv  \frac{T'}{T} ~~ ; ~~~~~~  
\beta\equiv\frac{\Omega'_{b}}{\Omega_{b}}
\ee
where $T$ ($T'$) is the ordinary (mirror) photon temperature 
in the present Universe,\footnote{The ratio of the temperatures 
between two sectors is nearly constant during the evolution of the Universe. 
In general, one has $T'/T = x [g_{\rm s}(T) / g'_{\rm s}(T')]^{1/3}$,
where the factors $g_{\rm s}$ and $g'_{\rm s}$ accounting for the 
degrees of freedom of the two sectors can be different 
from each other, see ref. \cite{z} for details.}  
and $\Omega_{b}$ ($\Omega'_{b}$) is the 
ordinary (mirror) baryon density fraction.  
In this section, we discuss the dependence of the length scales
relevant for structure formation from these parameters.

In the most general context, the present energy 
density contains relativistic (radiation) component 
$\Omega_r$, non-relativistic (matter) component 
$\Omega_m$ and the vacuum energy 
density $\Omega_\Lambda$ (cosmological term). 
The present observational data indicate that 
the Universe is almost flat, i.e.  
$\Omega_0=\Omega_m + \Omega_r + \Omega_\Lambda \approx 1$   
in a perfect accordance with the inflationary paradigm, 
with $\Omega_m = 0.2-0.3$ and the rest 
of the energy density is due to the cosmological term. 
In the context of our model, the relativistic fraction 
is represented by the ordinary and mirror 
photons and neutrinos,  
$\Omega_rh^2=4.2\times 10^{-5}(1+x^4)$, where
contribution of the M-species is negligible  
due the BBN constraint $x^4 \ll 1$. 
As for the non-relativistic component, 
it contains the O-baryon fraction $\Omega_b$ and
the M-baryon fraction $\Omega'_b = \beta\Omega_b$,     
while the other types of dark matter, e.g. the CDM, 
could also present and so in general 
$\Omega_m=\Omega_b+\Omega'_b+\Omega_{\rm CDM}$.\footnote{
In the context of supersymmetry,   
the CDM component could exist in the form of 
the lightest supersymmetric particle (LSP).  
It is interesting to remark that the mass fractions  
of the ordinary and mirror LSP are related as 
$\Omega'_{\rm LSP} \simeq x\Omega_{\rm LSP}$. 
In addition, an HDM component $\Omega_\nu$ 
could be due to neutrinos with mass $\leq 0.3$ eV. 
The contribution of the mirror neutrinos 
scales as $\Omega'_\nu = x^3 \Omega_\nu$ and thus   
it is irrelevant.}
In the following we use the central values by WMAP 
$\omega_b = \Omega_b h^2 \approx 0.023$ and   
$\omega_m = \Omega_m h^2 \approx 0.135$ \cite{wmap-data}, 
and consider scenarios with $\beta = 1 \div 5$ where the 
limiting case $\beta\simeq 5$ corresponds to the case
when the dark matter is entirely due to MBDM 
($\Omega_{\rm CDM}=0$).

The important moments for the structure formation  
are related to the matter-radiation equality (MRE)
and to the plasma recombination and matter-radiation 
decoupling (MRD) epochs. The MRE occurs at the redshift:   
\be \label{z-eq} 
1+z_{\rm eq}= \frac{\Omega_m}{\Omega_r} \approx 
\frac{3240}{1+x^4}\left(\frac{\omega_m}{0.135}\right) 
\ee
where we denote $\omega_m = \Omega_{m}h^2$. Therefore, 
for $x\ll 1$ it is not altered by the additional relativistic 
component of the M-sector.

The matter radiation decoupling takes place only after the most of 
electrons and protons recombine into neutral hydrogen   
and the free electron number density strongly diminishes,   
so that the photon scattering rate  
drops below the Hubble expansion rate. 
In the ordinary Universe the MRD takes place 
in the matter domination period, at the temperature 
$T_{\rm dec} \simeq 0.26$ eV which 
corresponds to redshift  
$1+z_{\rm dec}=T_{\rm dec}/T_{\rm today} \simeq 1100$. 

The MRD temperature in the M-sector $T'_{\rm dec}$ 
can be calculated following the same lines as in 
the ordinary one \cite{z}. 
Due to the fact that in either case the 
photon decoupling occurs when the exponential factor 
in the Saha equation becomes very small, 
we have $T'_{\rm dec} \simeq T_{\rm dec}$, 
up to small logarithmic corrections related to 
$\eta'$, different from $\eta$. Hence 
\be\label{z'_dec}
1+z'_{\rm dec} \simeq x^{-1} (1+z_{\rm dec}) 
\simeq \frac{1100}{x}\, 
\ee
so that the MRD in the M-sector occurs earlier than in the ordinary 
one. Moreover, for a value $x=x_{\rm eq}$, where: 
\be \label{x_eq} 
x_{\rm eq}=\frac{1+z_{\rm dec}}{1+z_{\rm eq}}
\simeq 0.34 \, \left(\frac{0.135}{\omega_{m}}\right) 
\ee
the mirror photon decoupling epoch coincides with the MRE 
epoch. This critical value plays an important role in our 
further considerations. Namely, for $x < x_{\rm eq}$ 
the mirror photons would decouple  
yet during the radiation dominated period. 

Let us discuss now the relevant scales for 
evolution of perturbations in the MBDM. 
The relevant scale for gravitational instabilities
is Jeans length, defined as the minimum scale at which, in the
matter dominated epoch, sub-horizon sized perturbations 
start to grow. The mirror Jeans scale is given by:
\be
\label{jeans}
\lambda'_{\rm J}(z) \simeq v'_s(z)\, (\pi/G\rho(z))^{1/2}\,(1+z)
\ee  
where $\rho(z)$ is the matter density at a given redshift 
$z$, $v'_s(z)$ is the sound speed in the M-plasma
and the $(1+z)$ factor translates the physical scale at the time
of redshift $(1+z)$ to the present scale.  
We remark that the M-plasma contains more baryons and less 
photons than the ordinary one, $\rho'_{b}=\beta \rho_{b}$ and 
$\rho'_\gamma = x^4\rho_\gamma$, and thus the sound speed
can have a quite different behaviour. We have:
\be \label{mirsound}
v'_s(z) \simeq 
{1 \over {\sqrt3}} \left(1+ {{3\rho'_b} \over {4\rho'_\gamma}}\right)^{-1/2}
\approx 
{1 \over {\sqrt3}} \left[ 1 +{3 \over 4} 
\frac{\beta \omega_b}{\omega_m}
(1+x^{-4}) \left({{1+z_{\rm eq}} 
\over {1+z}}\right)\right]^{-1/2}
\ee
where $\omega_b=\Omega_{b} h^2$,
quite in contrast with the ordinary world, 
where $v_s \approx c/\sqrt{3}$ practically 
till the photon decoupling.
The M-baryon Jeans length reaches the
maximal value at $z=z'_{\rm dec}$, where it is given by
\footnote{We assumed $\beta x^{-4} \gg 1$ so that the constant term in 
eq.~(\ref{mirsound}) can be neglected.}: 
\be
\lambda_{\rm J,dec}'\simeq
\frac{100\,x^{5/2}}{(\beta\omega_b)^{1/2}(x+x_{eq})^{1/2}} ~ {\rm Mpc} 
\ee
After decoupling, eq. (\ref{mirsound}) does not hold  
anymore and the Jeans scale decreases to very 
low values, due to the fact that the pressure supplied by the 
relativistic component of the mirror plasma disappears.

Density perturbations in MBDM on 
scales $\lambda\ge\lambda'_{\rm J,dec}$ which enter 
the horizon at $z\sim z_{\rm eq}$ undergo uninterrupted linear 
growth immediately after $z_{\rm eq}$. 
Perturbations on scales $\lambda\le\lambda'_{\rm J,dec}$ 
start instead to oscillate immediately after they enter the
horizon, thus delaying their growth till the epoch of
M-photon decoupling.

Finally, we turn our attention to dissipative processes which
can modify the purely gravitational evolution of perturbations.
As occurs for perturbations in the O-baryonic sector, 
also the M-baryon density fluctuations should undergo the 
strong collisional damping around the time of 
M-recombination. The photon diffusion from the overdense to underdense
regions induces a dragging of charged particles  
and washes out the perturbations at scales smaller than the 
mirror Silk scale, $\lambda'_S$.
The behavior of $\lambda'_S$ as a function of the parameter
$x$ and $\beta$ is given by
\be
\label{lambda_s}
\lambda'_S\simeq \frac{3\, f(x)}{(\beta\omega_b)^{3/4}} ~ {\rm Mpc} 
\ee
where $f(x)=x^{5/4}$ for $x > x_{\rm eq}$, 
and $f(x) = (x/x_{\rm eq})^{3/2} x_{\rm eq}^{5/4}$ 
for $x < x_{\rm eq}$. 

The impact of such scales on the evolution
of density perturbations will be discussed in the next section.

\section{Evolution of perturbations}

We clearly understand from the previous discussion
that MBDM has peculiar features which can leave a characteristic
imprint in the large scale structure of the Universe. 

{\it First}, perturbations in MBDM on scales 
$\lambda\le\lambda'_{\rm J,dec}$ experience an oscillatory regime. 
The MBDM oscillations transmitted via gravity to the ordinary baryons,  
could cause observable anomalies in LSS power spectrum 
and in the CMB angular power spectrum. 

{\it Second}, for $x\ge x_{\rm eq}$, the growth of perturbations on 
scales $\lambda\le\lambda'_{\rm J,dec}$ does not start at 
$z_{\rm eq}$ but is delayed till M-photon decoupling. 
If MBDM is the dominant dark matter component,
one expect to observe less structures on these scales than in
standard CDM scenario.

{\it Finally}, the density perturbation scales which can run 
the linear growth after the MRE epoch are limited by the 
length $\lambda'_S$. To some extent, the cutoff effect is  
analogous to the free streaming damping in the case of 
warm dark matter (WDM). 

In order to make quantitative predictions we computed 
numerically the evolution of adiabatic perturbations in a Universe in
which is present a significant fraction of mirror 
dark matter. More precisely, following the approach
described in \cite{mabert1}, we solved numerically in a 
synchronous gauge the linear evolution equations for perturbations 
of all matter components: ordinary baryons, photons, neutrinos,  
their mirror analogues, and cold dark matter.  
In fact, with respect to the standard case, the full set of equations 
was doubled in order to properly take into account the evolution of 
the mirror photon-baryon system. 
The decoupling in ordinary and mirror plasma was followed 
numerically as prescribed in \cite{mabert1}.
All computations were made assuming a flat space-time
geometry 
($\Omega_{0}=1$; i.e. $\Omega_{\Lambda}=1-\Omega_{ m}$).
In order to compare our predictions with the ``standard'' CDM results,
we have chosen a  ``reference cosmological model'' 
with scalar adiabatic perturbations and no massive neutrinos 
with the following set of parameters \cite{wmap-fit}:
\be\label{ref-model} 
\omega_{b} = 0.023, ~~ 
\omega_{m}= 0.133  ~~ (\Omega_{ m} = 0.25), ~~ 
\Omega_{\Lambda} = 0.75 , ~~  
n_{\rm s} = 0.97, ~~ h = 0.73 . 
\ee
We have included in this model the mirror sector and studied
the effects of MBDM as a function of the parameters $x$ and $\beta$.
For the sake of comparison, in all calculations the total amount of 
matter $\Omega_{m}=\Omega_{\rm CDM}+\Omega_{\rm b}+\Omega'_{\rm b}$
was maintained constant, $\Omega_m =0.25$. 
Mirror baryons contribution is thus always increased at the 
expenses of diminishing the CDM contribution. Hence, the case 
when dark matter is entirely represented by the MBDM, with 
no CDM component, corresponds to $\beta=\omega_m/\omega_b \simeq 5$.  

\begin{figure}[h]
\begin{center}
\leavevmode
{\hbox 
{\epsfysize = 5.3cm \epsffile{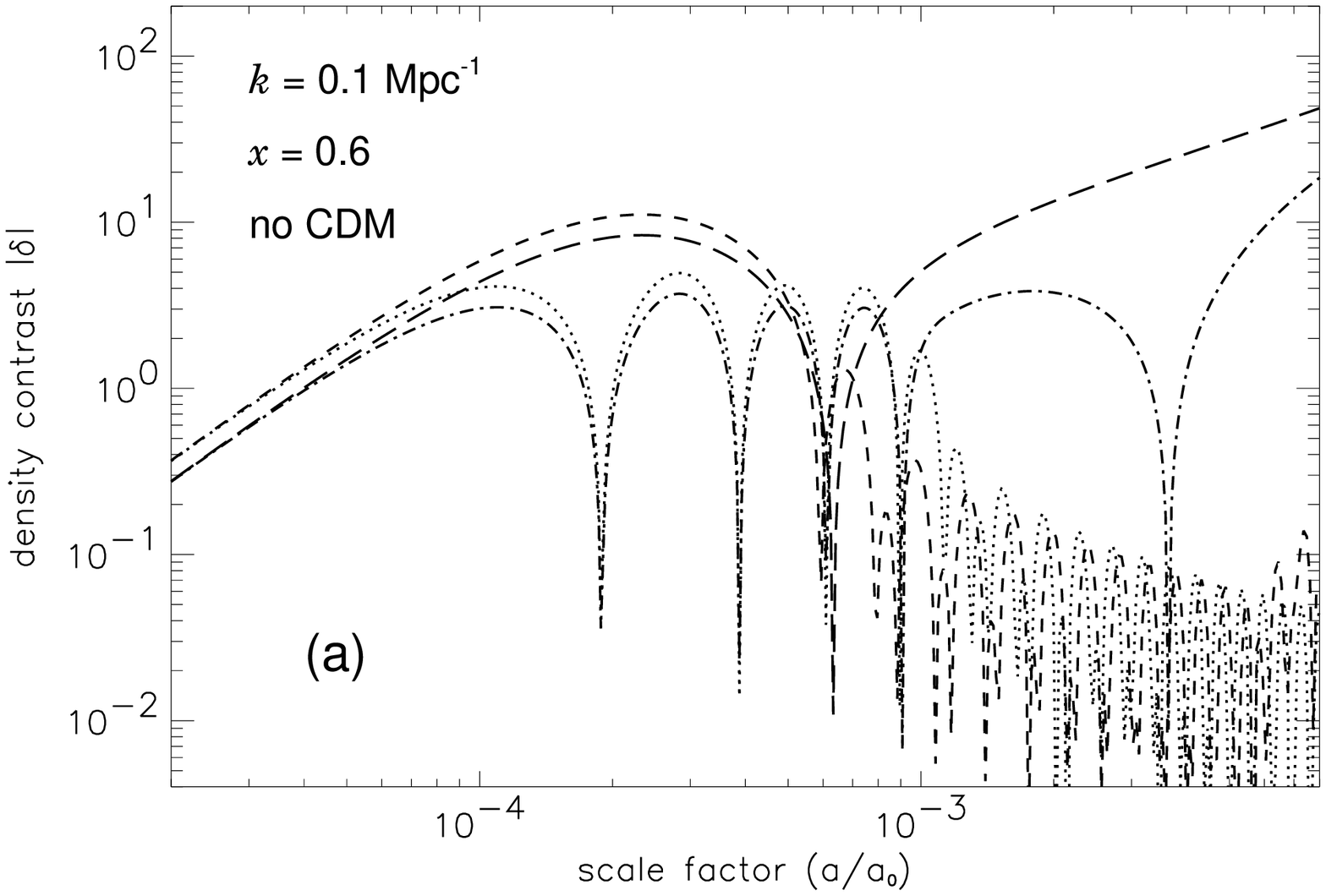} }
{\epsfysize = 5.3cm \epsffile{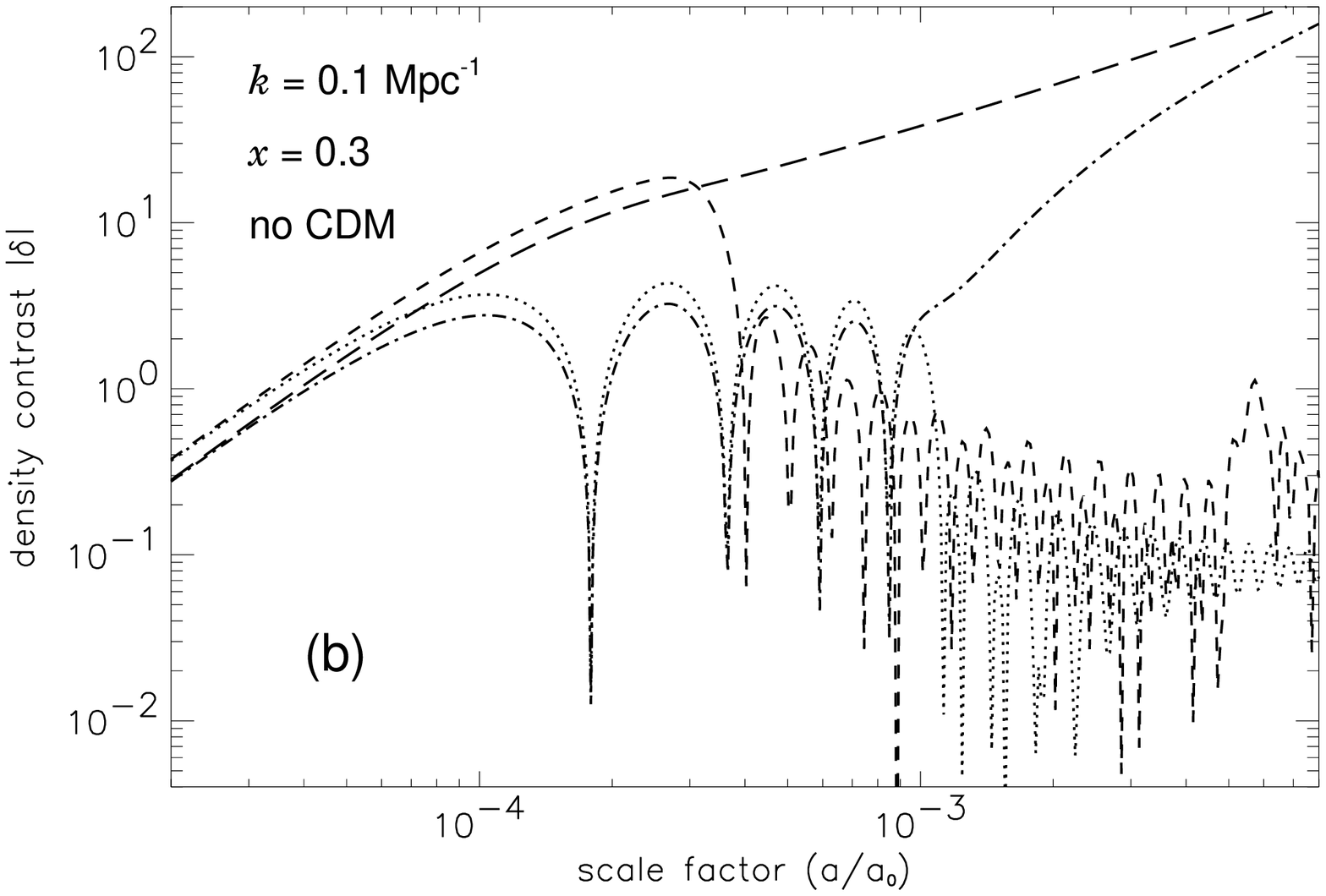} } }
\end{center}
\vspace{-.5cm}
\caption{\small Evolution of perturbations at the scale 
$k = 0.1$ Mpc$^{-1}$ in the case when dark matter is entirely due to
mirror baryons $(\Omega_{\rm CDM}=0)$. Dot-dashed  and dotted lines  
correspond to ordinary baryons and photons,  
while long dashed and dashed lines are for mirror baryons and photons. 
All cosmological parameters are taken as in (\ref{ref-model}). 
Left panel (a) corresponds to the case $x=0.6$ 
and the right panel (b) to the case $x=0.3$. }
\label{evolmirr}
\end{figure}

\begin{figure}[h]
\begin{center}
\leavevmode
{\hbox 
{\epsfysize = 5.3cm \epsffile{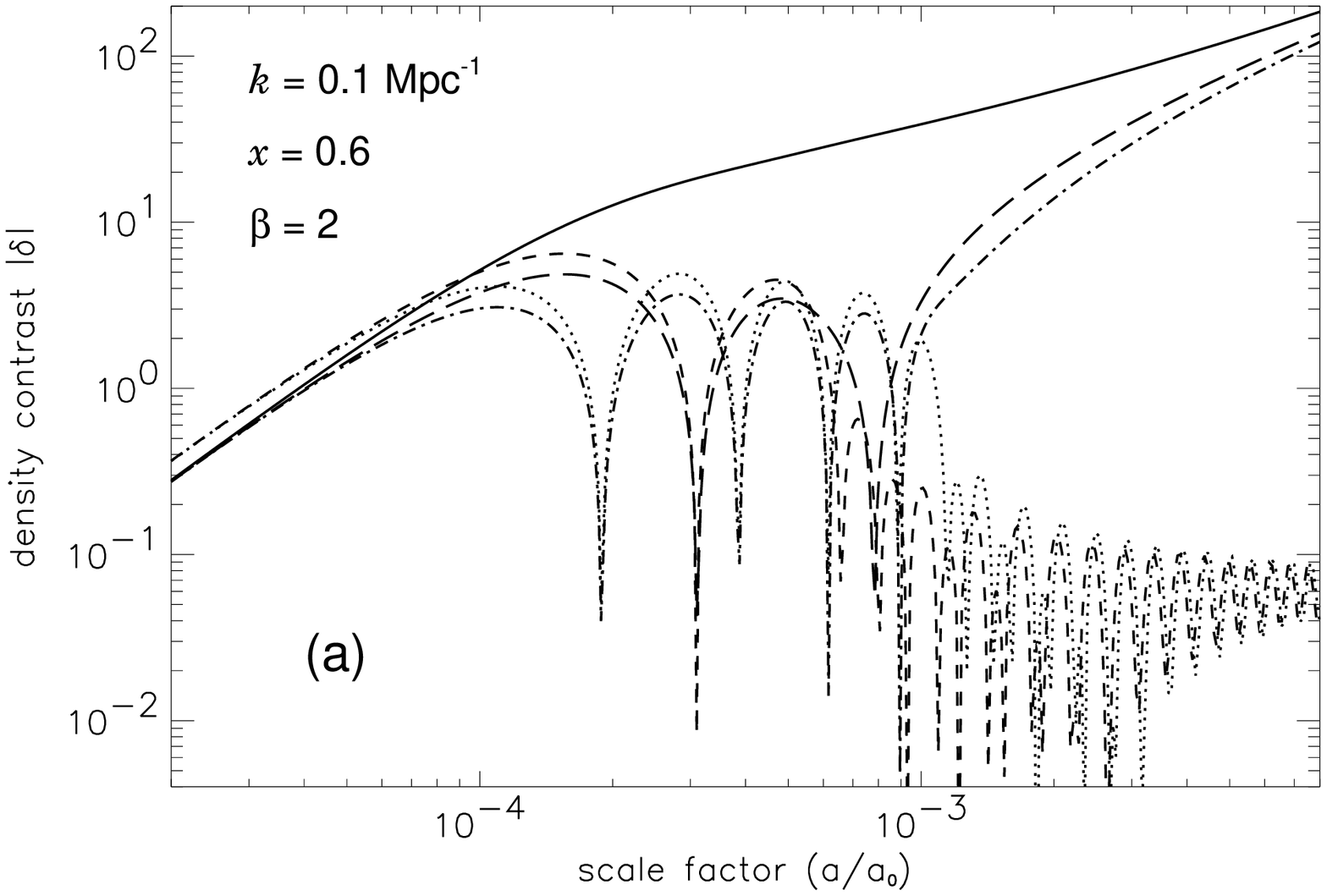} }
{\epsfysize = 5.3cm \epsffile{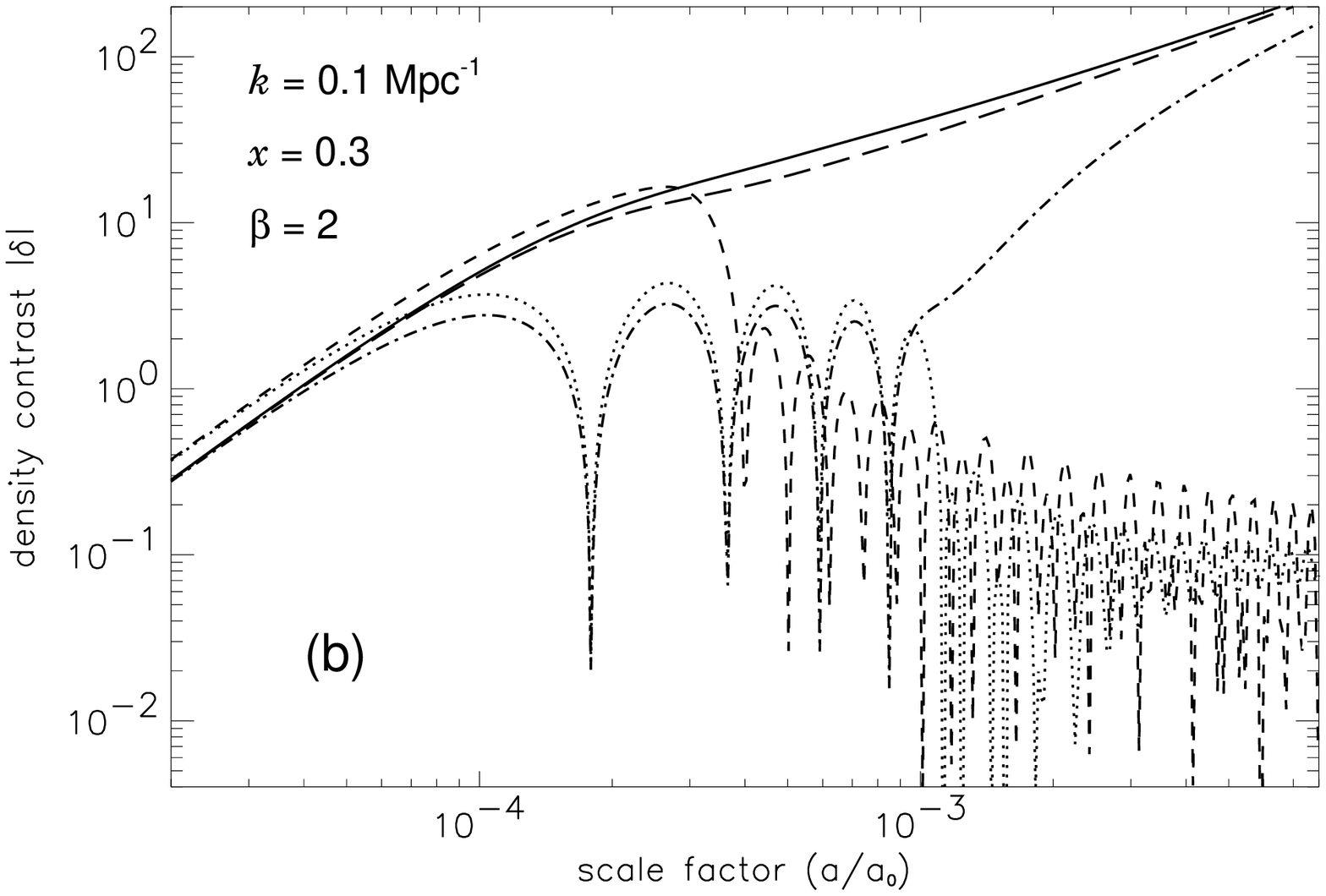} } }
\end{center}
\vspace{-.5cm}
\caption{\small The same as in Fig. \ref{evolmirr} but in the 
case when the MBDM is a sub-dominant dark matter component 
with $\beta=2$, i.e. $\Omega'_b=0.09$ and $\Omega_{\rm CDM}=0.12$.}
\label{evolcdm}
\end{figure}

To understand the impact of mirror matter 
on structure formation it is useful to look at 
Fig. 1 and Fig. 2 where, for a selected wavenumber 
$k = 2\pi/\lambda$ and for selected values of $x$ and $\beta$, 
we show the evolution of the density contrast in the various 
components, $\delta_i = \delta \rho_i/\rho_i$, 
as a function of the scale factor $a$.
In Fig. 1 we show the situation when 
dark matter is entirely due to mirror baryons, 
while in Fig. 2 mirror baryons represent a subdominant component.
The panels (a) of the two figures are obtained
by assuming large values of the mirror to ordinary temperature
ratio (i.e. $x=0.6$), while the panels (b) correspond to 
the value $x=0.3$ which, for the chosen cosmological parameters, 
implies that M-photon decoupling approximately coincides with 
MRE epoch (c.f. eq. (\ref{x_eq})).   

As long as the perturbation scale is larger than the horizon, 
the various components grow at the same rate ($\delta_i\propto a^{2}$).
The situation drastically changes when the perturbations enter the 
horizon (around $a/a_0 \sim 10^{-4}$). 
At this point, baryons and photons, 
for each sector separately, become causally connected 
and behave as a single fluid. In other words, we have two fluids, 
the ordinary baryon-photon and the mirror baryon-photon ones. 
The sub-horizon evolution of these 
fluids depends on the value of the Jeans lengths.
If the Jeans length is larger than the perturbation scale
then the  photo-baryon fluid starts to oscillate. This is always 
the case (before the decoupling) for ordinary baryons and photons.
This is not always true in the mirror sector.
By comparing panels (a) and (b) of these two figures we see in fact
that for large values of $x$ the M-photons and baryons 
oscillate, while for smaller $x$ values perturbations 
undergo uninterrupted growth even before M-photon decoupling 
(which occurs around $a/a_0\simeq 10^{-3}x^{-1}$). 

Oscillations in the mirror sector (when present),  
transmitted via gravity to the ordinary baryons,  
would produce observable anomalies in LSS power spectrum and
in the CMB anisotropy spectrum. By comparing the late time 
perturbation evolution in Fig. 1a and Fig. 2a, we can 
understand that the efficiency of this process depends on 
the amount of mirror matter present in the Universe. 
After decoupling, in fact, baryons, which are no longer
supported by photon pressure, rapidly fall in the 
potential wells created by the dominant dark matter component.
If M-baryons dominate the dark matter budget, this leads 
asymptotically to $\delta_b = \delta'_b$ 
(see late time evolution in Fig. 1a). 
This means that the baryonic structure power spectrum 
re-write the M-baryonic power spectrum, which
is suppressed at small wavelength due to Silk damping 
and is eventually modulated (if $x$ is not too small) 
as a results of acoustic oscillation.
If instead CDM is the dominant dark matter component,
we have asymptotically $\delta_b = \delta'_b = \delta_{\rm CDM}$
(see late time evolution in fig.2a), which means 
that both M-baryonic and O-baryonic structures 
will follow the ``standard'' CDM  power spectrum.    

The dependence of the  LSS power spectrum
on the parameters $x$ and $\beta$ is shown explicitly
in Fig. 3. In the upper panel we assume that the 
dark matter is entirely due to mirror baryons and we
consider variations of the $x$ parameter. 
For large $x$ values, as a result of the 
oscillations in MBDM perturbation evolution, one observes 
oscillations in the  LSS power spectrum.
The position of these oscillations depends on $x$, 
as can be easily understood. The smaller is $x$ the smaller 
is the mirror Jeans scale at decoupling and thus the smaller 
are the perturbations scales which undergo acoustic oscillations.
Superimposed to oscillations one clearly see  
the cut-off in the power spectrum due to the Silk damping.
We remark that the Silk scale $\lambda'_{s}$ also depends on $x$ 
(and $\beta$), as it is described by eq.~(\ref{lambda_s}).
As a consequence, the cut off in the power spectrum moves 
to smaller wavelength when we decrease the $x$ parameter.

In the lower panel of Fig. 3 we can appreciate the role of 
the parameter $\beta$. One clearly sees that, as expected, the
smallest is the amount of mirror baryons the less evident
are the features in the LSS power spectrum.  Interestingly, 
for large $x$ values, one observes a relevant effect even for 
relatively small amount of mirror baryons; even for $\beta=1$ 
the effects are quite noticeable. 

\label{lss_2}
\begin{figure}[h]
  \begin{center}
    \leavevmode
    \epsfxsize = 10cm
    \epsffile{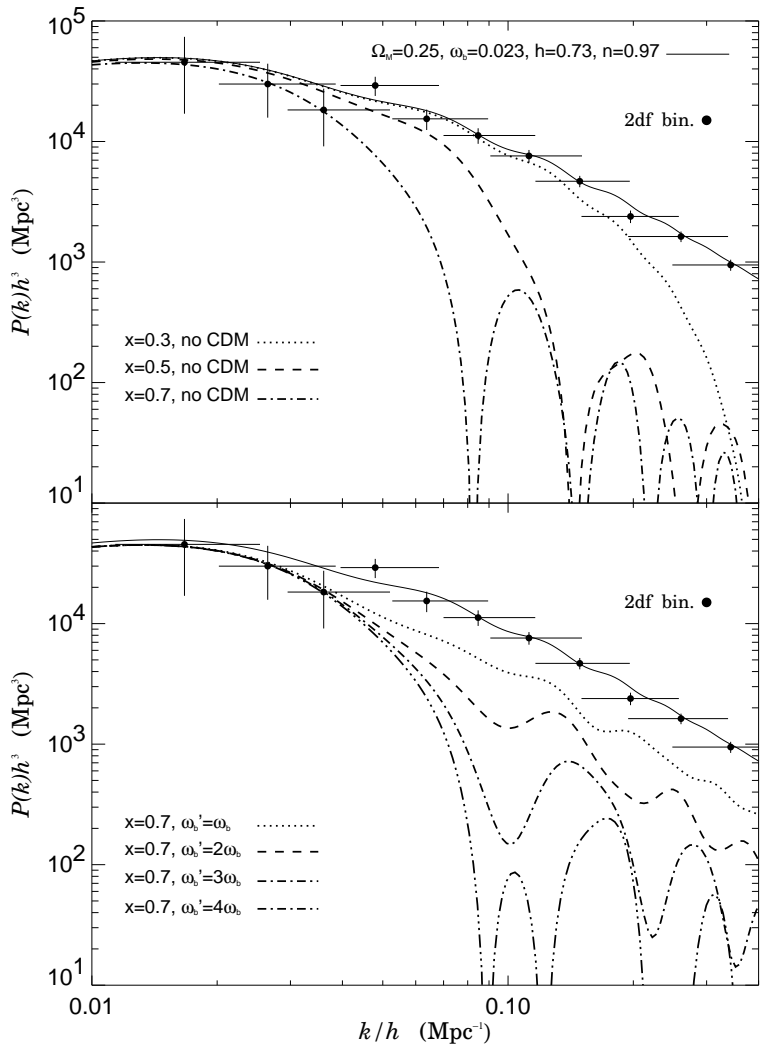}
  \end{center}
\vspace{-.4cm}
\caption{\small LSS power spectrum in the linear regime for 
different values of $x$ and $\omega_{\rm b}' \equiv \Omega_{\rm b}' h^2$,
 as compared with a standard model prediction
(solid line).
In order to remove the dependences of units on the Hubble constant, 
we plot on the x-axis the wave number in units of $ h $
and on the y-axis the power spectrum in units of $ h^{-3} $.
All parameters are taken as in (\ref{ref-model}). 
We also show the binned data of 2dF observations \cite{2df-teg}.
{\sl Top panel.} Models where dark matter is entirely due to MBDM  
(no CDM, i.e. $\beta=5$)
for different values $x = 0.3, 0.5, 0.7$. 
{\sl Bottom panel.} Models with mixed CDM+MBDM components,  
$\beta=1,2,3,4$ for a value of $x=0.7$. }
\label{cmblssfig3}
\end{figure}

Finally, in Fig. 4  we show the effects of MBDM on the angular spectrum 
of the CMB anisotropy. The predicted spectrum is quite strongly 
dependent on the value of $x$ (see upper panel), and it becomes 
practically indistinguishable from the CDM case for 
$x < x_{\rm eq}\simeq 0.3$, in this  case the MBDM 
behaves just as the CDM at the scales relevant for 
the CMB oscillations.  
However, the effects on the CMB spectrum rather weakly depend on the
fraction of mirror baryons (see lower panel for different values of $\beta$).  
In other words, the CMB anisotropy spectrum is mainly 
sensitive to amount of extra-radiation 
in the Universe due to the the mirror sector (which is fixed by $x$, 
being $\Omega'_r\propto x^4$) but it can hardly distinguish
between MBDM and CDM.

\begin{figure}[h]
  \begin{center}
    \leavevmode
   \epsfxsize =10cm
    \epsffile{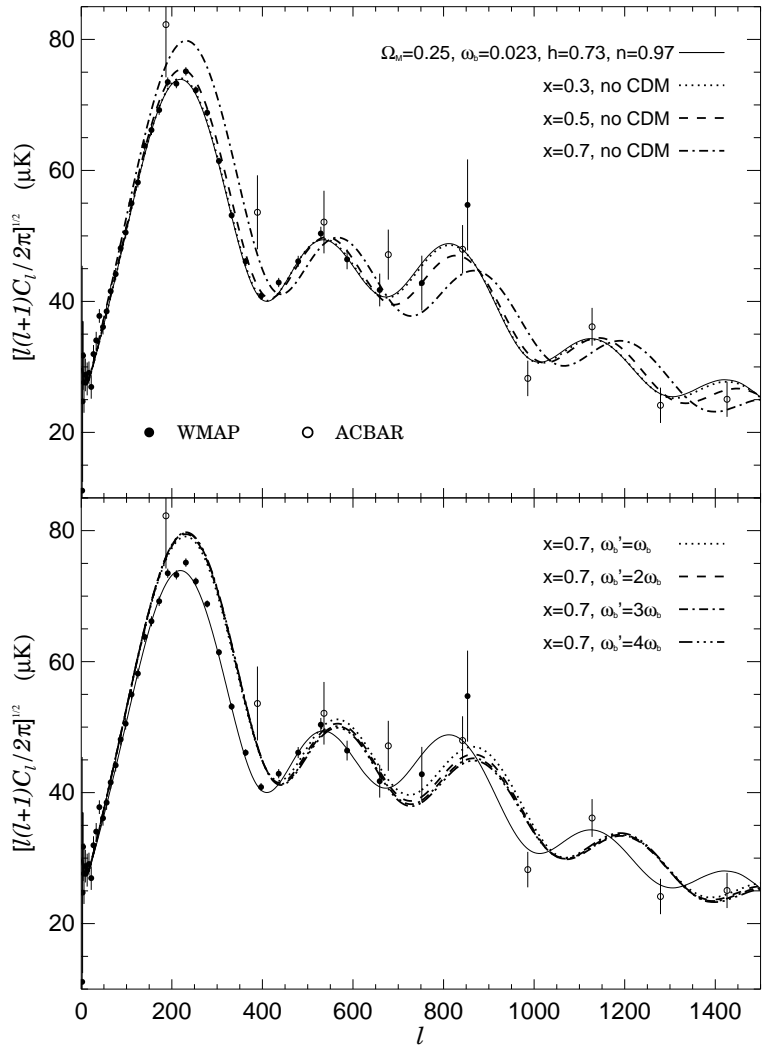}
  \end{center}
\vspace{-.4cm}
\caption{\small The CMB angular power spectrum for different values of 
$ x $ and $ \omega_{\rm b}'$, compared with 
a standard model (solid line). 
We also show the WMAP \cite{wmap-data} and ACBAR \cite{acbar-data} data.
The choices of the parameters for both {\sl top panel}
and {\sl bottom panel} exactly correspond to those of Fig. 3. }
\label{cmblssfig1}
\end{figure}
\vspace{.1cm}

Our predictions can be compared with the observational data
in order to obtain bounds on the possible existence of 
the mirror sector. To give a visual impression of the present 
situation we show in the Fig. 3 the 2dF binned data \cite{2df-teg}
and in Fig. 4 the WMAP \cite{wmap-data} and 
ACBAR \cite{acbar-data} data.

To extract information from the experimental data one clearly needs 
a detailed statistical analysis. However, some general conclusions 
can be obtained in a rather straightforward way: 
 
(i) The assumption that DM is entirely due to mirror baryons 
is evidently not compatible with present 
LSS data unless the value of $x$ is enough small: 
$x < x_{\rm eq} \approx 0.35$.

(ii) Very high values of $x$, $x > 0.5$, can be excluded
even for a relatively small amount of mirror baryons. 
E.g. for $x=0.7$, one has relevant 
effects on LSS and CMB power spectrum down to values of 
M-baryon density
of the order $\Omega'_{b} \sim \Omega_{b}$. 
On the other hand, intermediate values $x=0.3-0.5$ can be 
still allowed if the MBDM is a subdominant component of 
dark matter. 

(iii) For small values of $x$, say $x < 0.3$, 
neither the linear LSS power spectrum nor the CMB angular power 
spectrum can distinguish the MBDM from the CDM.   
In this case, in fact, the Jeans 
length $\lambda'_{\rm J,dec}$ and the Silk length $\lambda'_{S}$, 
which mark region of the spectrum below which one sees the effects of 
mirror baryons, decrease to very low values, which undergo 
non linear growth from relatively large red-shift.

\section{Conclusions}

The idea of a mirror sector of particles and interactions 
has attracted a significant interest over last years.  
The concept of mirror world could have interesting implications  
for the following problems: 
detection of Machos via gravitational microlensing \cite{BDG,Macho}, 
search ordinary-mirror star binaries \cite{Khlop} and mirror planetary 
objects \cite{planet}, 
implications of mirror matter for galaxy halos \cite{Halo},  
the role of the sterile neutrinos in neutrino physics \cite{FV-BM,neutrino} 
and their implications for ultra-high energy cosmic rays \cite{Venya} 
and gamma ray bursts \cite{GRB}, 
implications for the flavor and CP violation and axion physics 
\cite{PLB98,BGG}, etc. 
In particular, the effects based on the kinetic mixing among the 
ordinary and mirror photons \cite{Holdom,photon} can provide 
interesting possibilities for revealing the mirror matter in 
positronium decays \cite{Gninenko} and in dark matter 
detectors \cite{Foot}. 

In this paper, we have studied the cosmological implications
of the mirror matter as dark matter. 
More precisely, we have quantitatively discussed the 
effect of the MBDM on the evolution of density perturbations 
in the linear regime as function of the two free parameters in the model:
the temperature of the mirror plasma
(limited by BBN to $x = T'/T < 0.64$) 
and the amount of mirror baryonic matter 
$\beta = \Omega'_{b}/\Omega_{b} $.
We summarize here the main conclusions.

The concept of mirror baryons, as possible candidate for dark matter,
introduces two new scales in the 
structure formation scenario: the Jeans scale of the mirror 
photo-baryon fluid, $\lambda'_{\rm J}$, and 
the Silk damping scale of mirror baryons, $\lambda'_{\rm S}$. 
Due to the pressure support of mirror photons, perturbations in 
the mirror fluid on scales smaller than $\lambda'_{\rm J,dec}$ 
cannot grow before mirror photon decoupling. 
If decoupling occurs after the MRE epoch 
(i.e. if $x \ge x_{\rm eq}\simeq 0.35$),
and if mirror baryons are a relevant dark matter component, 
one expect then to see less structures on these scales with respect 
to the standard CDM scenario. 
In addition, mirror baryon perturbations on scales 
$\lambda\le\lambda'_{\rm J,dec}$ go through an oscillatory regime.
This could produce, via gravity, observable distortions in the CMB 
and LSS power spectrum. 
Finally, the mirror Silk scale $\lambda'_{\rm S}$
introduces a cut-off for the perturbation scales which can run 
the linear growth after matter-radiation equality.

All these effects were taken into account numerically in the Fortran 
code which we used to follow the perturbation evolution in presence 
of the mirror sector and to compute the LSS and CMB power spectra. 
The results are shown in Figs. 3 and 4 for various choices of 
parameters $x$ and $\beta$.
From a comparison with present observational data, 
one is able to conclude that the existence of a mirror sector with 
a relatively high temperature, $T' \geq 0.4 \;T$ 
and a high mirror baryon density $\Omega'_{b}\geq $ few $\Omega_{b}$ 
can be already excluded. 
This confirms previous estimates of refs. \cite{z,ignatiev} 
and improves the BBN bound on the mirror sector parameter space.
More details can be found in ref. \cite{Paolo}. 

In order to further reduce the allowed parameter space,
it would be clearly important to extend the analysis 
to the non-linear regime. 
Many relevant questions, related to the dynamics in this regime, 
can be formulated:
what implications could have the mirror Silk cutoff for scales which 
already went non-linear, e.g. for scanning the early Universe 
via studying reionization, quasars, Lyman-$\alpha$ forest, etc.?   
Can mirror baryons, being a dissipative dark matter component, 
provide extended triaxial halos instead of being clumped
into the galaxy as usual baryons?
How the  star formation mechanism proceeds in the M-sector
where the  temperature/density conditions and chemical contents
are much different from the ordinary ones?
Could the mirror protogalaxy at a certain moment before disk
formation become a collisionless system of the mirror stars and 
thus maintain a typical elliptical structure?
What is the initial mass function of mirror stars, and how many 
and how heavy mirror stars do we expect as Machos in the 
galactic halo? 

Many other questions can be formulated and various  new data 
are needed to discriminate different cosmological settings. 
Our numerical analysis, which describes the mirror world 
in the linear regime, 
sets the starting point for answering these questions.

\section*{Acknowledgements}

\noindent 
We thank Silvio Bonometto, Stefano Borgani, Alfonso Cavaliere, 
Andrei Doroshkevic and Alessandro Melchiorri for interesting discussions.
This work is partially supported by the MIUR research grant 
"Astroparticle Physics".

\end{document}